\documentstyle[12pt]{article}
\input mssymb                                      

\newfont{\twelvemsy}{msym10 scaled\magstep1}       
\newcommand{\real}{\mbox{\twelvemsy R}}            
\newcommand{\integer}{\mbox{\twelvemsy Z}}         

\def\thebibliography#1{{\noindent\bf References}\list
 {\arabic{enumi}.}{\settowidth\labelwidth{[#1]}\leftmargin\labelwidth
 \advance\leftmargin\labelsep
 \usecounter{enumi}}
 \def\newblock{\hskip .11em plus .33em minus .07em}
 \sloppy\clubpenalty4000\widowpenalty4000
 \sfcode`\.=1000\relax}

\newcommand{\rchi}{\raisebox{.4ex}{$\chi$}}

\begin{document}

\begin{flushright}
DAMTP R93/28 \\
hep-th/9310175 \\
October 1993
\end{flushright}
\vspace*{0.8cm}

\begin{flushleft}
{\Large\bf Spin Structures on Kleinian Manifolds}\\
\end{flushleft}

\begin{flushleft}
{\bf Lloyd Alty}\footnote{e--mail address: lja13{\char'100}amtp.cam.ac.uk}
{\bf and Andrew Chamblin}\footnote{e--mail address:
hac1002{\char'100}amtp.cam.ac.uk}

{\small Department of Applied Mathematics and Theoretical Physics,}\newline
{\small University of Cambridge, Cambridge CB3 9EW, England.}
\end{flushleft}
\vspace*{0.4cm}

{\noindent{\bf Abstract.} We derive the topological obstruction
to spin--Klein cobordism. This result has implications for signature change
in general relativity, and for the $N=2$ superstring.}
\vspace*{0.6cm}

{\noindent\bf 1.~ Introduction}

{\noindent In this paper we study topological obstructions to the global
existence of spin--Klein structure on four--manifolds. In particular, we wish
to determine the topological obstructions to spin--Klein cobordism (in analogy
with the recently found obstructions to spin--Lorentz cobordism, see
\cite{Cham1,GibbHawk,GibbHawk2}).

Let $M$ be any four--manifold. We will say that a metric on $M$ is of
{\it Kleinian signature} if it has signature $(--++)$, and we will say that $M$
admits {\it spin structure} if we are able to globally define a fibre bundle
over $M$ with structure group $Spin(2,2)$ (where $Spin(2,2)$ is the double
cover of $SO(2,2)$). It follows \cite{Cham2} that the obstruction to
spin structure is the vanishing of the second Stiefel--Whitney class $w_2$,
and we therefore need only understand the obstruction to putting a globally
non--singular Klein metric on $M$.

Let $\{\Sigma_1,\Sigma_2, \dots, \Sigma_n \}$ be an arbitrary collection of
three--manifolds, then a {\it cobordism} for
$\{\Sigma_1,\Sigma_2, \dots, \Sigma_n \}$ is a
four--manifold $M$ with $\partial M \cong \Sigma_1 \cup \Sigma_2 \cup \cdots
\cup \Sigma_n$ (where $\cup$ denotes disjoint union). We are interested in
cobordisms that possess more structure. In particular, we shall say that $M$ is
a {\it spin--Klein cobordism} for $\{ \Sigma_1,\Sigma_2, \dots, \Sigma_n\}$ if
and only if $M$ is a cobordism for $\{ \Sigma_1,\Sigma_2, \dots, \Sigma_n\}$,
and $M$ admits both spin structure and a globally defined non--singular metric
of Kleinian signature.

One motivation for this work is the recent interest in spacetimes in which
`signature change' is allowed \cite{Ellis,Hayward}. These signature change
studies have centered upon cosmological solutions to the classical
Einstein field equations in which the metric changes signature from $(++++)$
to $(-+++)$. It could be argued that this is too limiting, and more general
signature transitions, for example those from $(-+++)$ to $(--++)$ should be
considered \cite{Alty}. Our cobordism result is relevant to those cases in
which the spacetime metric changes to a Kleinian signature.

Our result may also prove to be of importance to the $N=2$ superstring. The
Weyl anomaly for this theory cancels provided that the string propagates in
a four dimensional target space. If the worldsheet is chosen to be of
Lorentzian signature then the target space must have Kleinian signature
\cite{Barrett}.}
\vspace*{0.4cm}

{\noindent\bf 2.~ Kleinian metric homotopy}

{\noindent In order to understand the topology of Kleinian metrics, we first
need to understand the obstruction to the existence of global Kleinian
structure. Indeed, we have from \cite{Steenrod}}\\
\vspace*{0.1cm}

{\noindent {\bf Lemma 1.} {\it Let $M$ be a smooth four--manifold without
boundary. Then $M$ admits a globally defined (non--singular) Kleinian
metric if and only if there exists a globally defined
(non--singular) field of 2--planes.}}\\ \vspace*{0.1cm}

A field of 2--planes may be defined in the following way.
Let $G_{2,4} \cong S^2 \times S^2$
denote the set of 2--planes in ${\real}^{4}$, then a
field of 2--planes is just a section of the fibre bundle which
has fibre $G_{2, 4}$.  If the section is everywhere non--vanishing then the
field of 2--planes is non--singular, otherwise the field of 2--planes is
singular. These definitions provide a  natural generalization of a
(non--singular/singular) vector field, which is a (non--vanishing/vanishing)
section of a sphere bundle. We will use the expressions `field of 2--planes'
and `plane field' interchangeably.

Each singularity of a singular plane field has
associated to it an index. Suppose that $x$ is a point in our
manifold $M$ at which some plane field $P$ becomes singular, and let
$S^{3}(x)$ denote a little sphere about $x$. Then the index indicates the
homotopy type of the map (defined by $P$) from $S^{3}(x)$ to  $G_{2,4}$.
However, such homotopy classes are in one--to--one correspondence with
elements of ${\pi}_{3}(G_{2, 4})~{\simeq}~ {\integer} ~{\oplus}~ {\integer}$.
Thus the index of $P$ at $x$ is classified by a pair of integers.

Let $M$ be an oriented compact four--manifold. Let $H$ denote the free abelian
group $H^2(M,\integer)/$torsion subgroup, and let $S$ denote the intersection
pairing on $H$ defined by the cup--product (ie. $S$ defines a map from
$H\otimes H$ to $\integer$ by taking the cup--product of elements in $H$ and
evaluating them on the fundamental orientation class of $M$). Define the coset
$W\subseteq H/2H$ by $w\in W$ if $S(w,x)=S(x,x)\;{\rm mod}\; 2$ for all $x\in
H$. Finally, let $\Omega$ denote the set of integers $\{S(w,w)| w\in W\}$.
We now recall an important theorem of Hirzebruch and Hopf \cite{HH} (also see
\cite{Thomas}).\\ \vspace*{0.1cm}

{\noindent {\bf Theorem 1.} {\it Let $M$ be an oriented compact four--manifold
without boundary. Then $M$ has a field of 2--planes with finite singularities.
The total index of such a field is given by a pair of integers $(a,b)$. The
following integers and only these, occur as the index for some plane field on
$M$:   \[
a =\frac{1}{4}\left(\alpha-3\sigma-2\rchi\right),\;
b = \frac{1}{4}\left(\beta-3\sigma+2\rchi\right)
\]
where $\alpha, \beta \in \Omega$, $\rchi=\rchi(M)$ denotes the Euler number of
$M$, and  $\sigma = \sigma(M)$ denotes the Hirzebruch signature of $M$.}}\\
\vspace*{0.1cm}

Now suppose we are given a cobordism $M$ for some collection of closed,
orientable three--manifolds $\{ \Sigma_1,\Sigma_2, \dots, \Sigma_n \}$.
Then we can always form the double of $M$, denoted $2M$, in the usual way
\cite{Spiv}.  The double has no boundary, and its Euler number is given by
$\rchi(2M) = 2\rchi(M)$. It also satisfies $\sigma(2M) = 0$ (since the two
`halves' of the double will have opposite orientations). Furthermore, we know
from \cite{HirzMay} that
\[
\alpha, \beta = \sigma(2M)\; {\rm mod}\; 8 \; .
\]
Combining these results with Theorem 1 we see that the total index
of any plane field on $2M$ has the following form:
\[
{\rm Total\; Index} = \left(2n-\rchi(M),2m+\rchi(M)\right)
\]
where $m,n \in {\integer}$.

Now push all of the singularities in the plane field on $2M$ over to `one of
the halves' of $2M$. Then we have constructed a non--singular
plane field on $M$ (by taking the singularity free half of $2M$), and by
construction, the degree of the map from  $\partial M$ to
$G_{2,4}$ (defined by the plane field) must be
\[
\left(2n-\rchi(M),2m+\rchi(M)\right) \;\;\;\;\;
{\rm for}\;\; m,n \in {\integer} \; .
\]
Recalling that a plane field defines a Klein metric up to homotopy, we shall
call this degree the {\it Klein kink} of the metric with respect to $\partial
M$ (in analogy with the Lorentz kink \cite{GibbHawk2}). Thus we have shown\\
\vspace*{0.1cm}

{\noindent {\bf Lemma 2.} {\it Let $M$ be any orientable manifold with
$\partial M \cong \Sigma_1 \cup \Sigma_2 \cup \cdots \cup \Sigma_n \neq
\emptyset$.  Then there are always globally defined non--singular plane
fields on $M$ and they must have Kleinian kink on $\partial M$ equal to
\[
\left(2n-\rchi(M),2m+\rchi(M)\right) \;\;\;\;\;
{\rm for}\; m,n \in {\integer} \; .
\] }}\\
\vspace*{0.1cm}

We can now combine Lemma 2 with the spin obstruction to get our main result.
\vspace*{0.4cm}

{\noindent\bf 3.~ Obstruction to spin--Klein cobordism}

{\noindent To begin, recall \cite{Cham1} the result that for manifolds $M$
with boundary we have the formula}
\[
(u(\partial M) + \rchi(M)) = \hat{w}_2\; {\rm mod}\; 2 \; ,
\]
where the mod 2 Kervaire semi--characteristic $u(\partial M)$ is defined as
\[
u(\partial M) = \dim_{{\Bbb Z}_2} \left( H_0(\partial M;{\integer}_2) \oplus
H_1(\partial M;{\integer}_2) \right)\; {\rm mod}\; 2 \; ,
\]
and $\hat{w}_2$ is defined in terms of the second Stiefel--Whitney class
$w_2$ as
\[
\hat{w}_2 = \left\{
\begin{array}{lrl}
0 & \;\;\; &
{\rm iff}\; w_2[c]=0\; {\rm for}\; {\rm all}\; {\rm 2\! -\! cycles}\;
c\in H_2(M) \\
1 & \;\;\; &
{\rm iff}\; w_2[c]=1\; {\rm for}\; {\rm some}\; {\rm 2\! -\! cycle}\;
c\in H_2(M)\; .
\end{array} \right.
\]

Now, suppose we are given a collection of three--manifolds
$\{\Sigma_1,\Sigma_2, \dots,\Sigma_n\}$ and a field of 2--planes (ie. a
Kleinian metric) on a collar neighbourhood of $\Sigma_i$. Then the field of
2--planes
will define a map from a collar neighbourhood of $\Sigma_1 \cup \Sigma_2 \cup
\cdots \cup \Sigma_n$ to $G_{2,4}$ and the homotopy type of the map  will be
indexed by the Klein kink, denoted  $kink(\Sigma_1 \cup \Sigma_2 \cup \cdots
\cup \Sigma_n; g_K)$, of the form  $(k_1,k_2) \in {\integer} \times {\integer}$
(we call $k_1$ and $k_2$ the first and the second component of the Klein kink
respectively).  We then have \\
\vspace*{0.1cm}

{\noindent\bf Theorem 2.} {\it Let $\{\Sigma_1,\Sigma_2, \dots,
\Sigma_n\}$ be some collection of closed orientable three--manifolds and let
$kink(\Sigma_1 \cup \Sigma_2 \cup \cdots \cup \Sigma_n;g_K)=(k_1,k_2)$ be the
degree of the map from $\Sigma_1 \cup \Sigma_2 \cup \cdots \cup \Sigma_n$ to
$G_{2,4}$ defined by some plane field on the collar neighbourhood of the
$\Sigma_i$. Then there exists a spin--Klein cobordism $M$ for the $\Sigma_i$
(in the sense that the degree of the map from $\partial M \rightarrow G_{2,4}$
is also $(k_1,k_2)$) if and only if
\[
\left( u(\partial M) + k_i \right) = 0\; {\rm mod}\; 2 \; ,
\]
where $k_i$ is either of $k_1$ or $k_2$}.\\
\vspace*{0.1cm}

{\noindent\it Proof.} Suppose such a spin--Klein cobordism $M$ exists.
Then $w_2 = 0$, and so
$u(\partial M)=\rchi(M)\; {\rm mod}\; 2$. Furthermore, by Lemma 2 we
know that
\[
kink(\partial M;g_K) = (2n-\rchi,2m+\rchi) \; .
\]
We see that the parity of $2n-\rchi$ or $2m-\rchi$ is determined solely by
$\rchi$  (ie. $k_i = \rchi\; {\rm mod}\; 2$), and
hence $\left( u(\partial M) + k_i \right) = 0\; {\rm mod}\; 2$.

Conversely, suppose $\left( u(\partial M) + k_i \right) = 0\; {\rm mod} 2$.
Take any globally defined Klein metric $g_K$ on $M$ then we have
\[
k_1 = k_2 = \rchi\; {\rm mod}\; 2 \; .
\]
Hence $\left( u(\partial M) + \rchi \right) = 0\; {\rm mod}\; 2$ and so
$w_2=0$. Thus $M$ is spin--Klein. $\blacksquare$
\vspace*{0.4cm}
\newpage

{\noindent\bf 4.~ Examples and Conclusions}

{\noindent Some examples of our obstruction to Spin--Klein cobordism are as
follows:}\\
\vspace*{0.1cm}

{\noindent\it Example 1.} Let $S^2$ be the two--sphere and $D^2$ a closed
two--disk, then we may put a Kleinian metric $g_K$ on the product manifold
$M\cong S^2 \times D^2$ by simply taking the product metric induced by a
signature $(++)$ metric on $S^2$ and a signature $(--)$ metric on $D^2$. The
boundary of $M$ is then given by $\partial M \cong S^1\times S^2$, and the
induced metric on  $\partial M$ is non--singular with signature $(-++)$.
Since the Klein kink of $g_K$ on $\partial M$ could also be
calculated using intersection theory (by counting the number of degenerate
points), it follows that $kink(\partial M; g_K) = (0,0)$. Hence, since $u(S^1
\times S^2) = 0$, we have that $S^2 \times D^2$ is a spin--Klein cobordism for
$S^1\times S^2$ with zero Klein kink.\\
\vspace*{0.1cm}

{\noindent\it Example 2.} Take ${\real}^4$ with ordinary flat metric $g_K$
of signature $(--++)$, and let $M\cong D^4$ be some closed $4$--ball in
${\real}^4$. Then $M$ is itself a Klein manifold with boundary $\partial M
\cong S^3$. In this case, we can clearly identify the plane field $P$
(corresponding to $g_K$) with a pair of non--vanishing vectors
$\{ v_1, v_2 \}$
which span $P$. In other words, the section $P$ of $G_{2,4}$ reduces to a
section of the Stiefel manifold $V_{2,4}$. It follows that
\[
kink(\partial M;g_K) =
\left( kink(\partial M;v_1), kink(\partial M;v_2) \right)
\]
where
$kink(\partial M;v_1) = kink(\partial M;v_2) = 1$ denote the kink of the
respective vector field on $\partial M$. Hence, since $u(S^3)=1$, we see that
$D^4$ is an example of a spin--Klein cobordism for $S^3$ with $kink(S^3;g_K)
= (1,1)$.\\
\vspace*{0.1cm}

{\noindent\it Example 3.} As was noted in the introduction, our result may be
applied to spacetimes that change signature from $(-+++)$ to $(--++)$.
Consider a spacetime
$M\cong M_L \cup M_K$ with $\partial M_L = \partial M_K = \Sigma
\neq \emptyset$, such that the metric $g_L$ on $M_L$ is of signature $(-+++)$
and the metric $g_K$ on $M_K$ is of signature $(--++)$. We require that the
two metrics agree on $\Sigma$, and hence the induced metric on $\Sigma$ is of
signature $(-++)$. Now, restricting our attention to $(M_K,g_K)$, it
follows that $kink(\partial M_K; g_K) = (0,0)$ and hence, from our theorem,
$M_K$ has spin structure if and only if $u(\partial M_K) = 0$. Thus, since
$u(S^3)=1$, we see that it is not possible for `bubbles' of Kleinian
signature to form across an $S^3$ surface whilst maintaining spin structure.\\
\vspace*{0.1cm}

Finally, we note that there is probably a generalization of this work to
non--orientable manifolds in the form of a topological obstruction to {\it
pin}--Klein cobordism (cf. \cite{Cham2}).
In analogy with the Lorentz case, this will presumably require us to pass to
non--oriented plane fields (ie. sections of the bundle of non--oriented
planes).

\vspace*{0.5cm}
{\noindent\it Acknowledgements.} The authors thank Gary Gibbons for
many helpful discussions.  A.C. is supported by NSF Graduate Fellowship
No. RCD--9255644.

\newpage

\end{document}